\documentclass[twocolumn,showpacs,preprintnumbers,amsmath,amssymb]{revtex4}

\usepackage{graphicx}
\usepackage{dcolumn}
\usepackage{bm}

\begin{document}


\title{Brillouin Scattering Studies of La$_{0.77}$Ca$_{0.23}$MnO$_3$ Across Metal-Insulator Transition}

\author{Md. Motin Seikh}
 \altaffiliation[Also at ]{Solid State Structural Chemistry Unit, Indian Insitute of Science, Bangalore 560 012, INDIA}
\affiliation{Chemistry and Physics of Materials Unit, Jawaharlal Nehru Centre for Advanced Scientific Research, Jakkur P.O., Bangalore 560 064, INDIA}
\author{Chandrabhas Narayana}
 \email{cbhas@jncasr.ac.in}
\affiliation{Chemistry and Physics of Materials Unit, Jawaharlal Nehru Centre for Advanced Scientific Research, Jakkur P.O., Bangalore 560 064, INDIA}
\author{L. Sudheendra}
\affiliation{Chemistry and Physics of Materials Unit, Jawaharlal Nehru Centre for Advanced Scientific Research, Jakkur P.O., Bangalore 560 064, INDIA}
\author{A. K. Sood}
 \altaffiliation[Also at ]{Department of Physics, Indian Insitute of Science, Bangalore 560 012, INDIA}
\affiliation{Chemistry and Physics of Materials Unit, Jawaharlal Nehru Centre for Advanced Scientific Research, Jakkur P.O., Bangalore 560 064, INDIA}
\author{C.N.R. Rao}
 \altaffiliation[Also at ]{Solid State Structural Chemistry Unit, Indian Insitute of Science, Bangalore 560 012, INDIA}
\affiliation{Chemistry and Physics of Materials Unit, Jawaharlal Nehru Centre for Advanced Scientific Research, Jakkur P.O., Bangalore 560 064, INDIA}

\date{\today}








\begin{abstract}
Temperature-dependent Brillouin scattering studies have been carried out on La$_{0.77}$Ca$_{0.23}$MnO$_3$ across the paramagnetic insulator - ferromagnetic metal (I-M) transition.  The spectra show a surface Rayleigh wave (SRW) and a high velocity pseudo surface acoustic wave (HVPSAW) besides bulk acoustic waves (BAW).  The Brillouin shifts associated with SRW and HVPSAW show blue-shifts, where as the frequencies of the BAW decrease below the I-M transition temperature (T$_C$) of 230 K.  These results can be understood based on the temperature dependence of the elastic constants.  We also observe a central peak whose width is maximum at T$_C$.
\end{abstract}


\maketitle


\section{Introduction}

The study of the perovskite manganites, A$_{1-x}$B$_x$MnO$_3$ (ABMO) where A and B are trivalent and divalent ions, respectively, has attracted much attention in recent years due to their fascinating properties and technological potential \cite{jin94}.  In particular, La$_{1-x}$Ca$_x$MnO$_3$ (LCMO) composition exhibit a transition from a paramagnetic insulating to a ferromagnetic metallic state for $0.2 \leq x \leq 0.5$ and show colossal magneto-resistance.  In ABMO type manganites, the transition temperature as well as other properties are markedly affected by the average radius of the A-site cations, which in turn affects the band width of the {\it e$_g$} electrons due to the Mn$^{3+}$ ions.  Jahn-Teller based electron-phonon coupling \cite{millis96,roder96} as well as the double exchange \cite{zener51} play important roles in determining the properties of the manganites.  There is experimental evidence for the presence of lattice distortions \cite{dai96,billinge96,radaelli96,zhao96} which give rise to polarons which may be of lattice and/or of magnetic origin \cite{teresa97,erwin96}. 

Brillouin scattering is a powerful probe to study the surface and bulk accoustic phonons as well as magnetic excitiations in opaque solids \cite{murugavel00}. The surface phonons can give rise to a surface Rayleigh wave (SRW), a pseudo surface acoustic wave (PSAW) and a high velocity pseudo surface acoustic wave (HVPSAW) \cite{carlotti92}.  The Poynting vector of SRW lies parallel to the free surface and its particle displacement field decays exponentially within the medium.  On the other hand, both PSAW and HVPSAW radiate energy into the bulk and get attenuated due to their decay into bulk phonons.  Unlike the SRW, the PSAW propagates only along some specific directions of the surface of anisotropic media and its phase velocity is $\sim$ 40 \% higher than the SRW velocities \cite{cunha95,cunha98,cunha01}.  In comparison, the HVPSAW propagates in both anisotropic as well as isotropic media \cite{camley85} and has a phase velocity nearly twice that of regular SRW \cite{cunha95,cunha98,cunha01}.  Both the PSAW and the HVPSAW have potential applications in surface accoustic wave devices due to their higher frequencies as compared to that of SRW \cite{cunha95,cunha98,cunha01}.

We have carried out Brillouin scattering study of La$_{0.77}$Ca$_{0.23}$MnO$_3$ (LCMO23), which exhibits the paramagnetic insulator - ferromagnetic metal (I-M) transition at T$_C =$ 230 K.  In addition to the SRW, the HVPSAW modes are observed for the first time in this manganite.  The SRW and HVPSAW mode frequencies increase where as the frequencies of the bulk phonons decrease below T$_C$.  This contrasting temperature-dependence can be understood in terms of the temperature-dependence of the elastic constants.  A central peak, whose width is maximum at T$_C$, is also observed and we also examine the origin and behavior of the central peak in this paper.

\section{Experimental Details}

Polycrystalline powders of LCMO23 were prepared by the solid-state reaction of stoichiometric amounts of lanthanum acetate, calcium carbonate and manganese dioxide. The materials were ground and heated at 1000 $^\circ$C for 60 hrs with two intermediate grindings. The sample was then further heated at 1200 $^\circ$C for 48 hrs. The polycrystalline powder was filled in a latex tube and pressed using a hydrostatic press at a pressure of 5 tons. The rod thus obtained was sintered at 1400 $^\circ$C for 24 hrs. The rod was then used to grow these crystals by the floating zone melting technique.  The technique employs SC-M35HD double reflector infrared image furnace (Nichiden Machinery Ltd., Japan). The growth rate and rotation speeds were 10 mm/h and 30 rpm, respectively.  The composition of the crystal was determined using Energy Dispersive Analysis of X-rays (EDAX) using LEICA S440I scanning electron microscope (SEM) of M/s LEICA, Japan using a Si-Li detector. The percentage of Mn$^{+4}$ was determined to be 23 $\pm$ 2 \%.  Figure 1 shows the resistivity and d.c. magnetization of the sample.  The peak in the resistivity at T$_C$ = 230 K marks the I-M transition.

\begin{figure}
\includegraphics{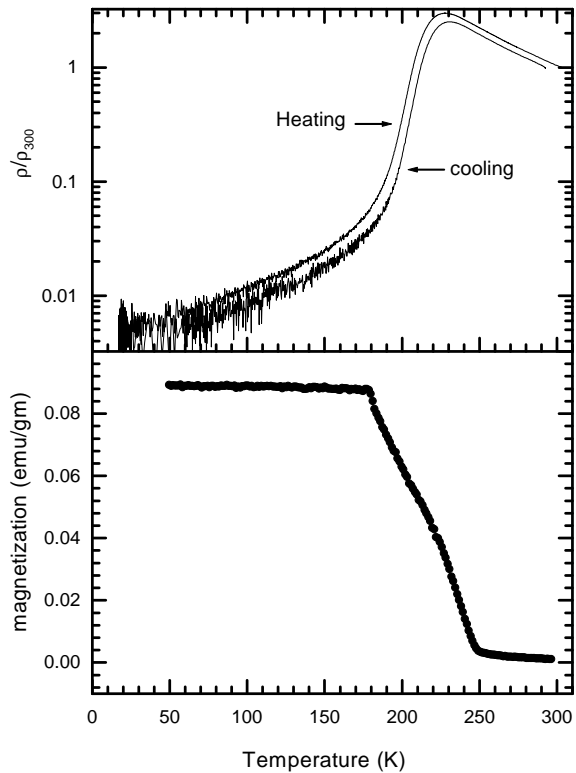}
\caption{\label{fig1} Temperature dependence of (a) Resistivity ($\rho/\rho_{300}$), where $\rho_{300}$ is the resistivity at 300 K (b) Magnetization of LCMO23.}
\end{figure}

Brillouin spectra were recorded in the back-scattering geometry with an incident angle $\theta = 45^\circ$ with the surface normal using a JRS Scientific Instruments 3+3 pass tandem Febry-Perot interferometer equipped with a photo avalanche diode as a detector.   The Nd:YAG single mode solid-state diode-pump frequency doubled laser (Model DPSS 532-400, Coherent Inc. USA) operating at 532 nm and power of $\sim$ 25 mW (focused to a diameter of $\sim$ 30 $\mu$m) was used to excite the spectra. The temperature experiments were carried out using a closed cycle helium cryostat (CTI Cryogenics, USA). The sample temperature was measured within an accuracy of $\pm$ 1 K.  There  is no laser damage of the sample surface as verified under a high magnification microscope.  The typical time required per spectrum was 1.5 hrs.

The laser induced heating of the sample was estimated to be 66 K as measured from the anti-Stokes to Stokes intensity ratio of the 480 cm$^{-1}$ Raman band of the same sample recorded using similar experimental conditions. The data presented here have been corrected to take into account the laser heating.  

\section{Results and Discussion}

Figure 2 shows the Brillouin spectra recorded using two free spectral ranges.  The spectra reveal four modes: 6.8 GHz (labeled as R$_1$), 15.3 GHz (R$_2$), 24.1 GHz (B$_1$) and 58.2 GHz (B$_2$).  Figure 3 shows the linear dependence of the R$_1$ and R$_2$ mode frequencies on the magnitude of the wave vector parallel to the surface, $\vec{q}_\parallel$, thereby establishing them to be the surface modes.  Upon rotating the crystal about the normal to the surface, the R$_1$ mode frequency shows an oscillatory behavior as a function of the rotation angle $\phi$, as expected for a SRW. The R$_2$ mode is associated with the HVPSAW, since its frequency is almost double than that of the R$_1$ mode with much broader line width, similar mode has been observed in GaAs along (1$\bar{1}$1) \cite{carlotti92} and discussed in the literature \cite{cunha95,cunha98,cunha01}.  In addition, a weak but sharp peak is found at 8.6 GHz.  This mode appears at certain orientations of the crystal and its frequency mode also shows a linear dependence on $\vec{q}$$_\parallel$, suggesting this to be the PSAW mode.  The PSAW and HVPSAW modes appear in our spectra because the sample surface is not along any specific crystallographic direction. The peaks labeled B$_1$ and B$_2$ in Fig. 2 are associated with the bulk acoustic waves (BAW).  The line widths of these modes are much higher due to the uncertainity in the wave-vector arising from the finite penetration depth of the incident radiation.

\begin{figure}
\includegraphics{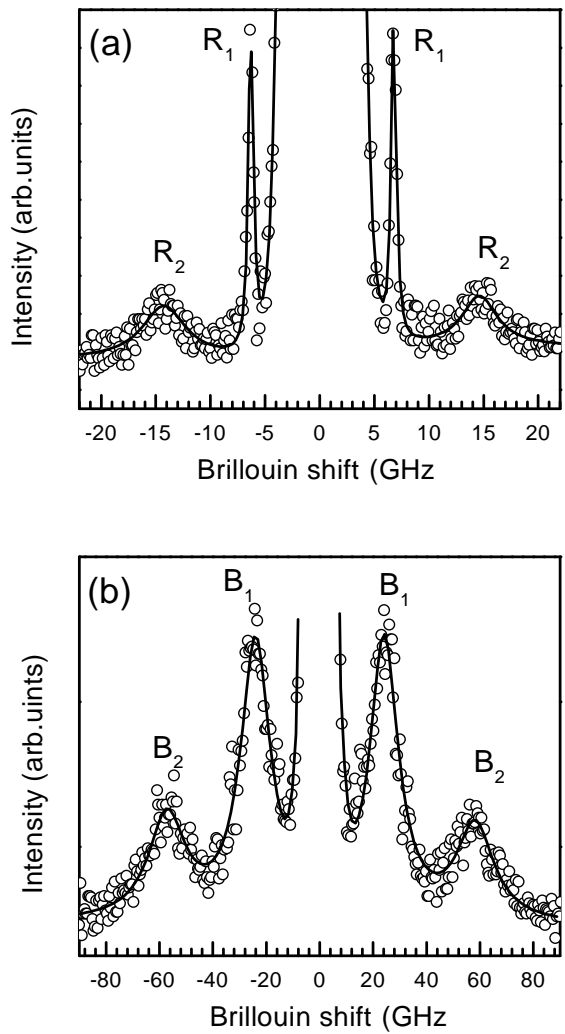}
\caption{\label{fig2}Room temperature spectrum of LCMO23 (a)recorded with free spectral range (FSR) 25 GHz (b) FSR 85 GHz. The solid lines through the data points are fits to the data, using two lorentzian peaks and an appropriate background on both Stokes and anti-Stokes side.}
\end{figure}

\begin{figure}
\includegraphics{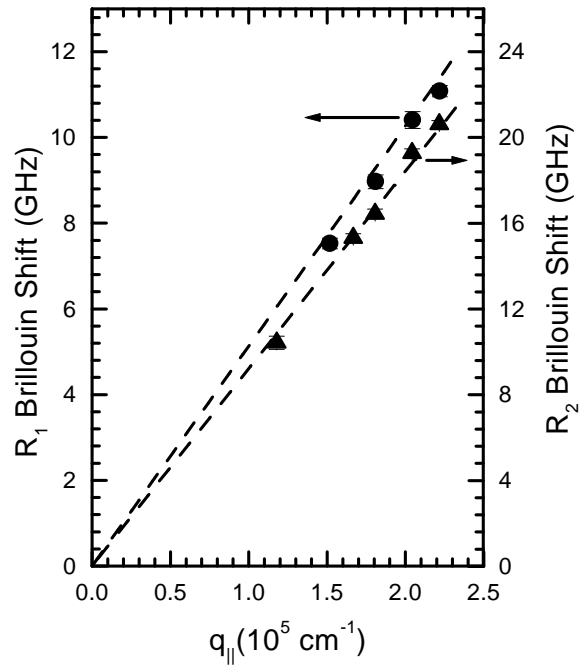}
\caption{\label{fig3} The frequency dependence of both R and R$_\circ$ mode as a function of $\vec{q}_\parallel$.  The solid line is a linear fit to the data points shown in solid circles and triangles.  The slope of the line is the surface velocity.}
\end{figure}

The temperature dependence of the changes in frequencies of all the four modes is shown in Fig. 4, where $\Delta \omega = \omega(T)- \omega(370 K)$.  We see from Fig. 4(a) that the frequency of the SRW (R$_1$) mode increases below T$_C$.  There is a gradual increase in the HVPSAW mode frequency as the temperature is lowered.  The B$_1$ and B$_2$ modes, on the other hand, show a downward jump in frequency at T$_C$. 

\begin{figure}
\includegraphics{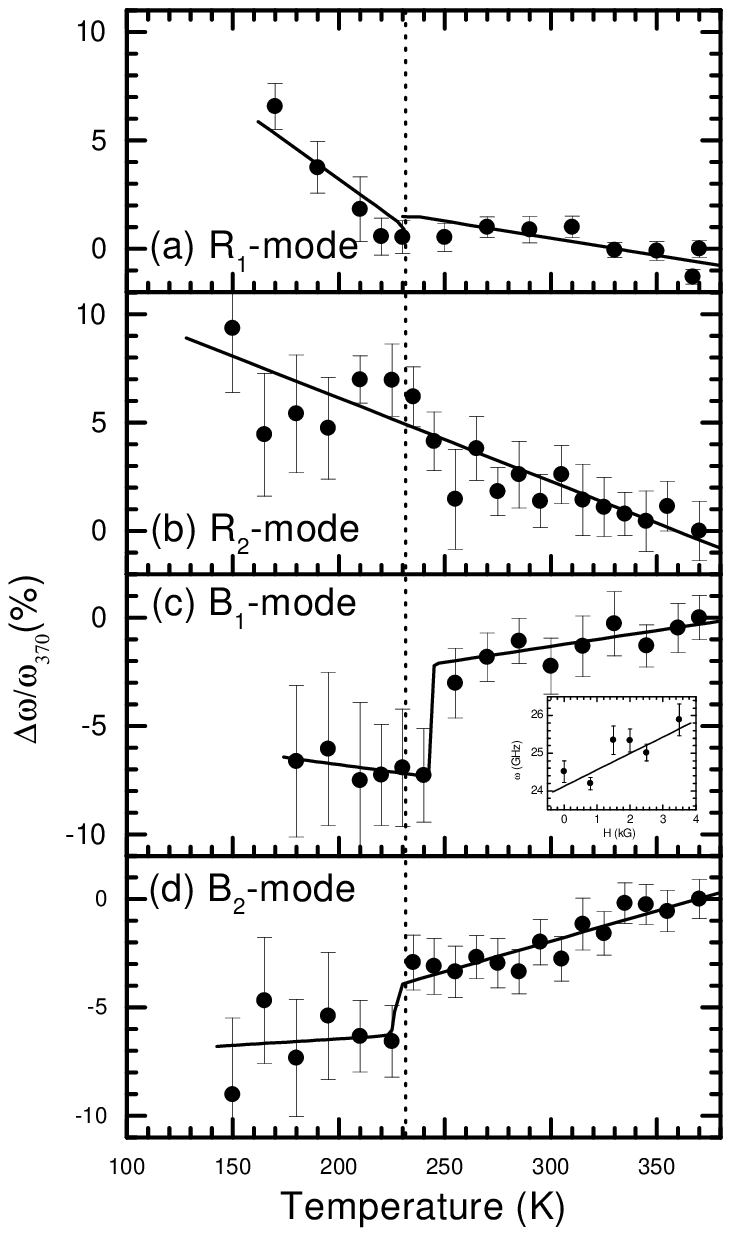}
\caption{\label{fig4} Percentage change in frequency as a function of temperature (a) R mode, (b) R$_\circ$ mode, (c) B$_1$ mode and (d) B$_2$ mode.  Inset in (c) shows the magnetic field dependence of B$_1$ mode.  The lines drawn are linear fits to act as guide to the eye.}
\end{figure}

We now discuss the temperature dependence of the various modes.  The SRW frequency can be calculated using the Greens function \cite{cummins01,zhang98}, which requires the values of the elastic constants.  Since the orthorhombic distortions in LCMO23 are not large, we can assume it to be cubic and use the three elastic constants C$_{11}$, C$_{12}$ and C$_{44}$  to calculate the SRW frequency.  Unfortunately, the elastic constants of LCMO are not available in the literature and we have, therefore, used the experimental values of elastic constants of La$_{0.835}$Sr$_{0.165}$MnO$_3$ (LSMO165) \cite{hazama00}, measured as a function of temperature using ultrasonic attenuation.  The elastic constant data are reproduced in Figure 5(a).  The Greens function used here to calculate the SRW frequency of LSMO165 is for the (100) surface.  The elastic constant values of LCMO23 should be similar to LSMO165, since the calculated SRW mode frequency for LSMO165 for $\theta = 45^{\circ}$ matches the value observed experimently in LCMO23 at 300 K.  

\begin{figure}
\includegraphics{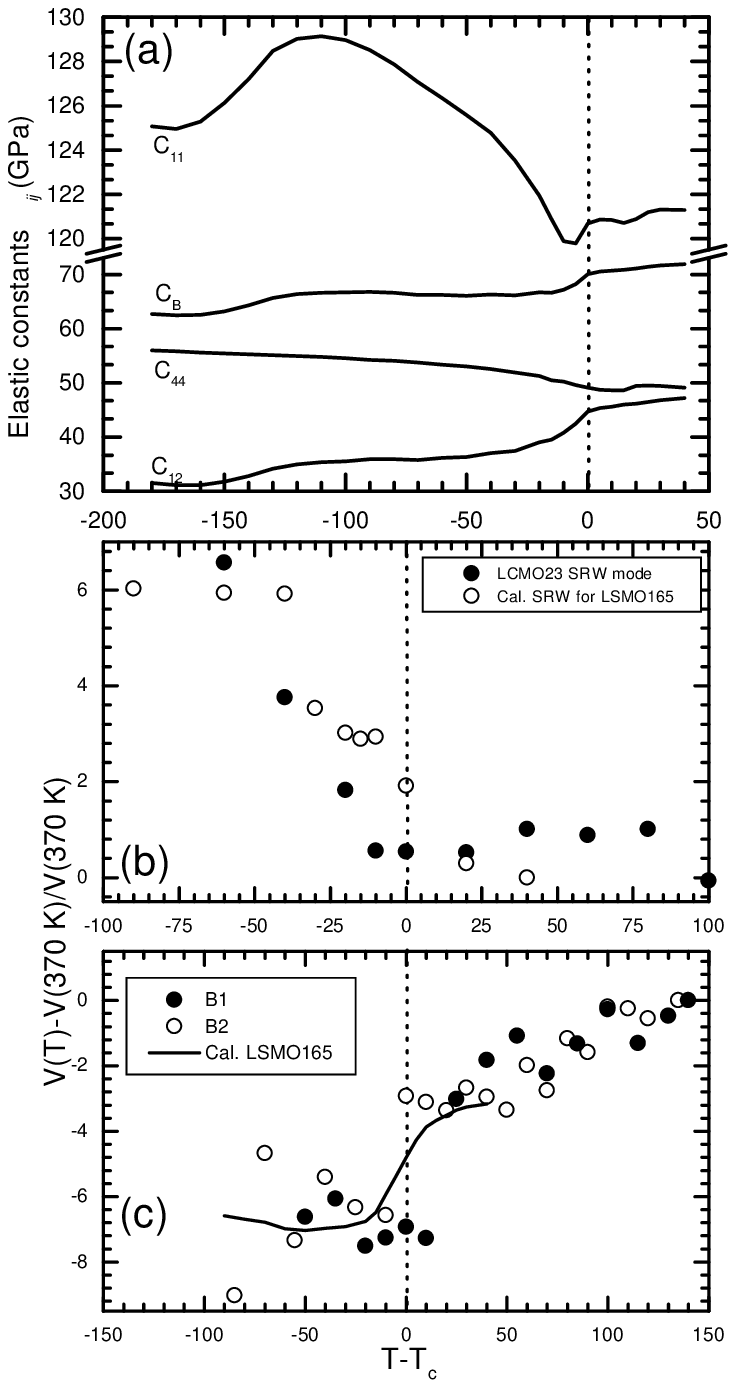}
\caption{\label{fig5}(a) The temperature dependence of C$_{11}$, C$_{44}$, C$_{12}$ and C$_B$ for LSMO165 reproduced from Ref. [19]. (b)  Percentage change in surface sound velocity of LCMO23 and LSMO165 (calculated) as a function of reduced temperature (T$_C -$T). (c) Percentage change in bulk sound velocity of LCMO23 of both B$_1$ (solid circles) and B$_2$ (open circles) modes and LSMO165 (calculated using C$_B$ values in (a) (solid curve)) as a function of reduced temperature (T$_C -$T).}
\end{figure}

Figure 5(b) shows the relative change in the surfrace Rayleigh sound velocity as a function of the reduced temperature ($T - T_C$) for LCMO23 (experimental) as well as LSMO165 (calculated).  It is seen that the there is a hardening of the SRW mode velocity below the T$_C$ and that the relative change in the SRW mode velocity is similar in LSMO165 and LCMO23.  In  La$_{1-x}$Sr$_x$MnO$_3$ (LSMO), hardening of bulk sound velocity is observed, for 0.11 $\leq x \leq$ 0.17, below the T$_C$ \cite{fujishiro97}.  A 5 \% hardening in the bulk sound velocity is found in the case of La$_{0.67}$Ca$_{0.33}$MnO$_3$ below the T$_C$ \cite{ramirez96}.  The origin of the hardening of the sound velocity below T$_C$ should be the same in all these materials exhibiting an I-M transition.  It is suggested \cite{lee97} that the mode frequency increases below T$_C$ due to the decrease in the electron-phonon coupling arising from the high mobility of carriers in the metallic state.  Application of an external magnetic field further increases the sound velocity, since the magnetic field greatly enhances the mobility of the carriers \cite{lee97}.  In the high temperature insulating phase, the electron-phonon coupling is larger, giving rise to a smaller frequency of the accoustic phonon.  The temperature dependence of the HVPSAW mode frequency, also arises due to the increase in C$_{11}$ and C$_{44}$ across the I-M transition, just as in the case of the SRW mode.  

We have carried out the experiments in the 90$^\circ$ geometry, in the case of the B$_1$ mode, to avoid strong reflection artefacts peaks, which appear around 27 and 33 GHz in the back-scattering geometry. The inset in Fig. 4(c) displays the magnetic field ($H$) dependence of the B$_1$ mode frequency showing an increase in the frequency with the magnetic field.  In general, the magnetic exciation with frequency $\omega$ depends on $H$ as $\omega = \Delta + Dq^2 + \gamma H$, where $\Delta$ is the anisotropic spin wave energy gap, $D$ is spin stiffness coefficient, $q$ is wavevector transfer and $\gamma$ is the gyromagnetic ratio.  This means, if the B$_1$ mode were a magnetic excitation, the slope (see inset of Fig. 4(c)) should be equal to $\gamma$.  The observed slope is much smaller than the expected value of $\gamma$ (typically $\sim$ 2), suggesting that the B$_1$ mode is not a magnetic excitation.  It is therefore a BAW with a strong magneto-elastic coupling.  The mode B$_2$ does not show any dependence on H.

A $\sim$ 3 \% softening is observed across the I-M transition temperature in the B$_1$ and B$_2$ modes (cf. Fig. 4(c) and (d)).  This is in contrast to the behavior of the surface phonon and bulk sound velocities found in the ultrasonic measurements \cite{fujishiro97,ramirez96,lee97}.  Taking into account the temperature dependence of elastic constants in Fig. 5(a) and considering that the bulk accoustic mode in a general direction can be expressed in terms of the bulk modulus, C$_B$ (= (C$_{11} +$ 2C$_{12}$)/3) \cite{hazama00}, we have calculated the relative change in the sound velocity ($v$) of the LSMO165 using the equation $v = \sqrt{\frac{C_B}{\rho}}$.  In Fig. 5(c), we have overlayed this plot on the relative changes in the sound velocity for B$_1$ and B$_2$ modes of LCMO23 as a function of reduced temperature ($T - T_C$).  It is interesting that the percentage change in the velocities are similar in the experimentally observed data of LCMO23 and the calculated data of LSMO165.  Since our experiments were carried out on an arbitrary cut crystal, the B$_1$ and B$_2$ modes should be related to the bulk modulus changes.  The present study suggests that the decrease in C$_{12}$ and hence in C$_B$ below the T$_C$ is responsible for the temperature dependence of the B$_1$ and B$_2$ modes.  

The intensity of the R$_2$ and B$_2$ modes do not show considerable changes across the I-M transtion.  On the other hand, the intensity of the R$_1$ mode decreases through out the temperature range, making it difficult to measure it below 100 K.  The B$_1$ mode also cannot be observed below 140 K.  The linewidths of both the surface phonons do not change as a function of temperature, where as the linewidths of the bulk phonons increase with decreasing temperature as shown in Fig. 6.  This can arise from the increase in the uncertainity in  the wavevector transfer due to the reduced penetration depth of light with the increase in electronic conductivity \cite{sandercock82}.

\begin{figure}
\includegraphics{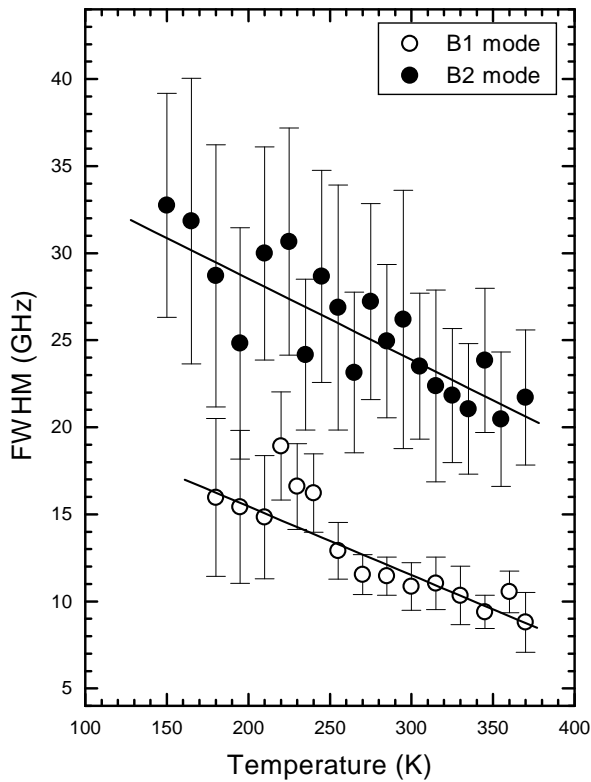}
\caption{\label{fig6} The temperature dependence of the full-width at half-maxima (FWHM) for the B$_1$ (open circles) and B$_2$ (solid circles) modes.  The solid lines are linear fit to the data.}
\end{figure}

In Fig. 7, we show the Brillouin spectra at various temperatures around T$_C$, in the 90$^\circ$ scattering geometry, revealing the presence of a central mode between 290 K and 180 K.  We used three Lorentzian peaks, with an appropriate linear background, to fit the spectrum at each temperature.  For temperatures below 180 K and above 290 K, we obtained a perfect fit (as shown in panels (a) and (b) of Fig. 7) with the elastically scattered mode giving an FWHM of 0.6 GHz over the entire temperature range of 80 to 370 K (except at low temperatures where the B$_1$ mode disappears, only one Lorentzian with a linear background was used to fit the spectra, as in the case of Fig. 7(e)).   Between 290 K and 180 K, it was difficult to fit the spectra in this manner.  Taking the FWHM of the elastically scattered mode as 0.6 GHz and adding an additional Lorentzian centered at $\omega = 0$, a good fit would be obtained as shown in (c) and (d) of Fig. 7.  This suggests the presence of a central mode in the temperature 290 - 180 K range.  Figure 8 shows the temperature dependence of both the FWHM and the intensity of the central mode, revealing that these quantities peak near the T$_C$.  A similar observation of the central peak has been made in neutron scattering experiments across the paramagnetic to ferromagnetic transition \cite{lynn96,dai00,dai01,bersuker90,baca98} at $q \sim 0.1$ to 0.3 \AA$^{-1}$ in La$_{1-x}$Ca$_x$MnO$_3$ ($x$ = 0.2, 0.25, 0.3 and 0.33), Nd$_{0.7}$Sr$_{0.3}$MnO$_3$ and Pr$_{0.63}$Sr$_{0.37}$MnO$_3$.  The presence of the central peak in the manganite suggests the presence of competing magnetic states. In neutron diffraction, the spectral weight of the spin-wave excitation starts to decrease as T $\rightarrow$ T$_C$, while the weight for the spin-diffusion component increases rapidly. This spin-diffusion component dominates the fluctuation spectrum near T$_C$, in marked contrast to conventional ferromagnets.  If one assumes the spin diffusion as the origin of the central peak in the present study, then the intrinsic width (FWHM or $\Gamma$) of the central peak is given by $\Gamma = \Lambda$q$^2$, where $\Lambda$ is the spin diffusivity.  In the present study, $\Gamma/2 \pi = 21$ GHz at T$_c$ (see Fig. 8(a)) and $\vec{q}$ = $3.675\times10^{-3}$ \AA$^{-1}$  leading to a value of spin diffusivity $\Lambda$ of 40.4 eV/\AA$^2$.  This value of $\Lambda$ is very much larger than  that observed in the case of Nd$_{1-x}$Sr$_x$MnO$_3$ ($x = 0.3$), where $\Lambda = 26(2)$ meV/\AA$^2$ \cite{baca98}.  Thus the origin of central peak in the present case does not appear to be spin diffusion.

\begin{figure}
\includegraphics{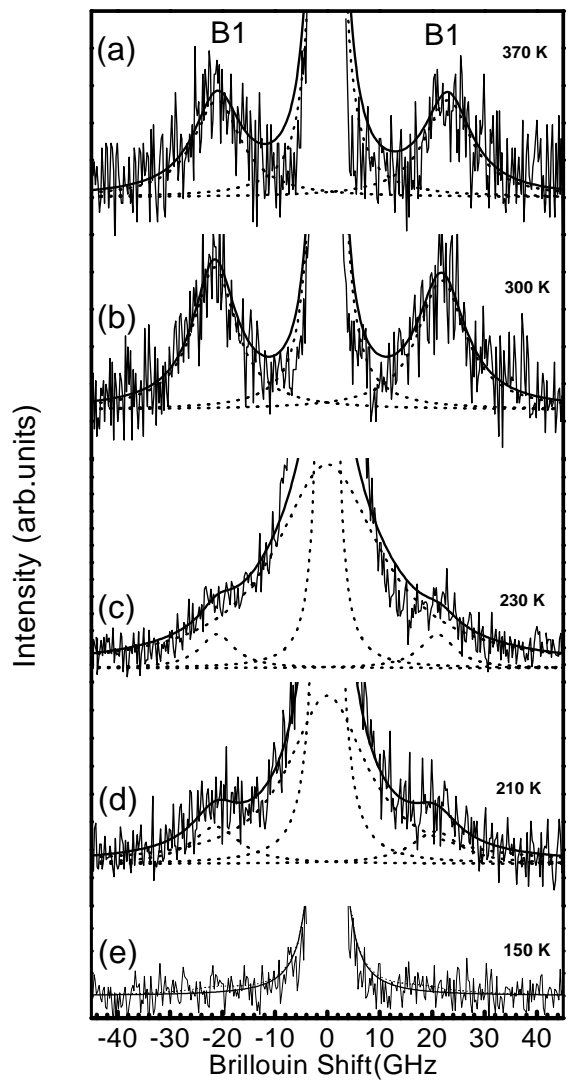}
\caption{\label{fig7} The Brillouin spectrum recorded at different temperatures showing the B$_1$ mode and the Rayleigh (elastic scattering) mode.  The solid curve represents the fit to the data using a set of Lorentzian peaks and an appropriate linear background.  The dotted curves represent the individual Lorentzians with the added linear background.}
\end{figure}

\begin{figure}
\includegraphics{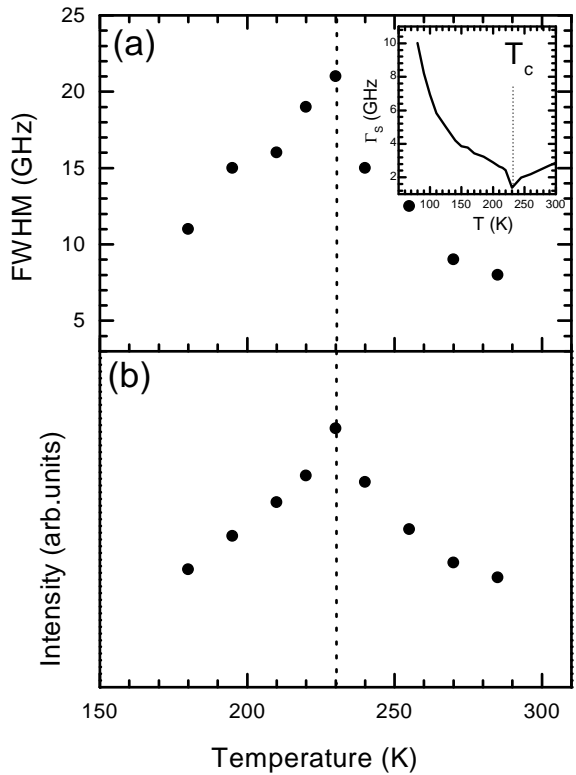}
\caption{\label{fig8}Temperature dependence of the central mode (a)FWHM and (b)intensity as deduced from the fits shown in Fig. 7.  The verticle dashed line represents the transition temperature (T$_C$).  Inset in (a) shows the calculated linewidth $\Gamma_S = 2 \kappa$q$^2$ as described in the text.}
\end{figure}

Entropy fluctuations with a width $\Gamma_S = 2 \kappa$q$^2$ \cite{sandercock82,fabelinskii94}, where $\kappa$ is the thermal diffusivity ($\kappa = \frac{k}{C \/\rho}$, where $k$ is the thermal conductivity, $C$ is the specific heat and $\rho$ is the density of the medium) can also be the origin of the quasielastic scattering.  The value of $k$ at 300 K is 22.9 mWcm$^{-1}$K$^{-1}$ (for LCMO (x=0.25)) \cite{fujishiro01}, while $C$ = 75.5 Jmol$^{-1}$K$^{-1}$ and $\rho$ = 6.2 g\/cm$^{-3}$ giving $\kappa = 1.1\times10^{-2}$ cm$^2$s$^{-1}$.  The width $\Gamma$ turns out to be 3 GHz.  Taking $\rho$(T), $C$(T) and $k$(T) for LCMO (x=0.25), the temperature dependence of $\Gamma_S$ is obtained as shown in the inset of Fig. 8. The temperature dependence  of $\Gamma_S$ is in contrast to that observed in Fig. 8(a) showing that entropy fluctuations are not responsible for the central mode.

Entropy fluctuations can also be viewed as fluctuations in the density of the phonon $\vec{q}$$_1$, which scatters the light by the two-phonon difference process invloving emission of a phonon $\vec{q}$$_1$ with the simultaneous absorption of a phonon $\vec{q}$$_1 - \vec{q}$ from the same phonon branch \cite{sandercock82}.  In the microscopic picture, scattering of light describes either the indirect \cite{landau34} or direct \cite{wehner72} coupling of light to phonon-density fluctuations, assuming at all times local thermodynamic equilibrium.  Along with these, there is also the contribution to phonon density fluctuations from additional ``dielectric fluctuations'', away from the local thermodynamic equilibrium \cite{coombs73}.  In this case, the two phonon difference process is dominated by zone boundary phonons where the density of states is high, leading to a broad central peak extending beyond the Brillouin peaks.  The line width of such a broad central peak is independent of q \cite{sandercock82}.  The central peak will, therefore, have a narrow component with a linewidth of $\kappa q^2$ and a q-independent broad component from "dielectric fluctuations". Since the central mode in our experiments, shown in Fig. 8, extends right upto the bulk modes, we suggest that it is associated with the q-independent "dielectric fluctuation".  It has been proposed that `defects' can contribute to the central peak \cite{schwabl91}.  Such defects can be the ferromagnetic  insulating phase present before the I-M transition in LCMO23 \cite{huang00}.

In conclusion, Brillouin scattering experiments on LCMO23 reveal, for the first time, the presence of a high velocity pseudo surface acoustic wave in manganites.  The hardening of the SAW frequency below T$_C$ arises from the temperature dependence of C$_{11}$ and C$_{44}$, where as the softening of the bulk phonons below T$_C$ can be attributed to the decrease in C$_{12}$ with the decreasing temperature.  The central peak is suggested to arise from non-equilibrium dielectric fluctuations  Linked to the presence of a ferromagnetic insulating phase preceeding the I-M transition.

AKS thanks Prof. J. D. Comins for the computer program to calculate the surface mode spectra using Greens function.  We would like thank Ms. G. Kavitha for implimenting the program and making the preliminary runs.  Mr. Seikh would like to thank Council of Scientific and Industrial Research (CSIR), India for a research fellowship.

\end{document}